\documentclass[twocolumn,prl,amsmath,letterpaper,floatfix,showpacs]{revtex4}
\usepackage{graphicx}
\usepackage{amsmath}
\usepackage{amssymb}
\usepackage{color}
\usepackage{units}

\begin{document}

\title{Quantum resonances in selective rotational excitation of molecules with a sequence of ultrashort laser pulses}

\author{S.~Zhdanovich$^{1}$, C.~Bloomquist$^{1}$, J.~Flo{\ss}$^{2}$, I.Sh.~Averbukh$^{2}$, J.W.~Hepburn$^{1}$, and V.~ Milner$^{1}$}

\affiliation{$^{1}$Department of  Physics \& Astronomy and The Laboratory for Advanced Spectroscopy and Imaging Research (LASIR), The University of British Columbia, Vancouver, Canada \\
$^{2}$Department of Chemical Physics, The Weizmann Institute of Science, Rehovot, Israel}

\begin{abstract}{We investigate experimentally the effect of quantum resonance in the rotational excitation of the simplest quantum rotor - a diatomic molecule. By using the techniques of high-resolution femtosecond pulse shaping and rotational state-resolved detection, we measure directly the amount of energy absorbed by molecules interacting with a periodic train of laser pulses, and study its dependence on the train period. We show that the energy transfer is significantly enhanced at quantum resonance, and use this effect for demonstrating selective rotational excitation of two nitrogen isotopologues, $ ^{14}N_2$ and $ ^{15}N_2$. Moreover, by tuning the period of the pulse train in the vicinity of a fractional quantum resonance, we achieve  spin-selective rotational excitation of para- and ortho-isomers of $ ^{15}N_2$.}
\end{abstract}


\maketitle

The periodically kicked rotor is a paradigm system for studying classical and quantum chaos \cite{Casati79}. In the quantum regime, the dynamics of the kicked rotor exhibits such fundamental phenomena as suppression of classical chaos \cite{Casati79}, Anderson localization in angular momentum \cite{Fishman82} and quantum resonances in the accumulation of rotational energy \cite{Izrailev80}. Even though these effects have been thoroughly studied with ultracold atoms in optical fields \cite{Moore95} and Rydberg atoms in microwave fields \cite{Koch95}, they have never been observed experimentally in a real rotational system.

It has been recently indicated \cite{Floss11} that quantum resonances can be detected in a system of true quantum rotors - an ensemble of diatomic molecules subject to periodic rotational kicking by ultrashort non-resonant laser pulses. It has been also proposed to use this quantum effect for selective laser manipulation of molecular mixtures \cite{Floss11}. Here, we demonstrate for the first time the fundamental phenomenon of quantum resonance in a real rotational system by studying the transfer of energy from a femtosecond pulse train to the rotation of different isotopologues of molecular nitrogen. We start with rotationally cold molecules and measure the energy transfer directly by means of a state-resolved detection of the rotational population in the excited molecular ensemble. To generate a train of ultrashort pulses, we employ a high-resolution femtosecond pulse shaper which enables us to scan the train period on the revival time scale. In our experiments, quantum resonance is manifested by a strong dependence of the population distribution width and the acquired rotational energy on the pulse train period with a sharp maximum at the rotational revival time.

Following the proposal of \cite{Floss11}, we also demonstrate isotope-selective rotational control in a mixture of $ ^{14}N_2$ and $ ^{15}N_2$ by tuning the train period near the quantum resonance, and spin-selective control in a mixture of different spin isomers of $ ^{15}N_2$ by kicking the molecules near the fractional quantum resonance.

Strong laser fields affect molecular rotation by exerting an angle dependent torque on the field induced molecular dipole \cite{Zon75, Friedrich95, Seideman95, Rosca02, Dooley03, Underwood05}. In the limit of ultrashort laser pulses, the pulse acts on a molecule as an instantaneous rotational ``$\delta $-kick''. In the quantum picture, a laser kick induces multiple Raman transitions between the rotational states of the molecule, transferring population from lower to higher rotational states and creating a coherent rotational wavepacket - a quantum-mechanical analogue of the ensemble of classical rotors. Because of the discrete energy spectrum of the rotational wavepacket, its dynamics is periodic in time, with a period known as the rotational revival time \cite{Robinett04}. Rotational revivals have been thoroughly studied in the context of molecular alignment \cite{Rosca02, Dooley03} - the appearance of a preferential direction in the distribution of molecular axes \cite{Stapelfeldt2003}.

Quantum periodicity of the rotational dynamics raises the question about the ability to enhance, and possibly control, molecular rotation with a periodic sequence of laser pulses (a ``pulse train'') \cite{Leibscher2003}. An accumulative effect of a pulse train on the degree of molecular alignment has indeed been observed when a train period exactly matched the revival time \cite{Cryan2009}, yet the direct measurement of the degree of rotational excitation and its dependence on the train period have not been studied.  Here, we detect the total energy transfer between the optical field and the molecules, showing that the accumulation of rotational energy in the molecular species resonant with the applied pulse train is substantially higher than in the off-resonance molecules.

\begin{figure}
\centering
    \includegraphics[width=1.0\columnwidth]{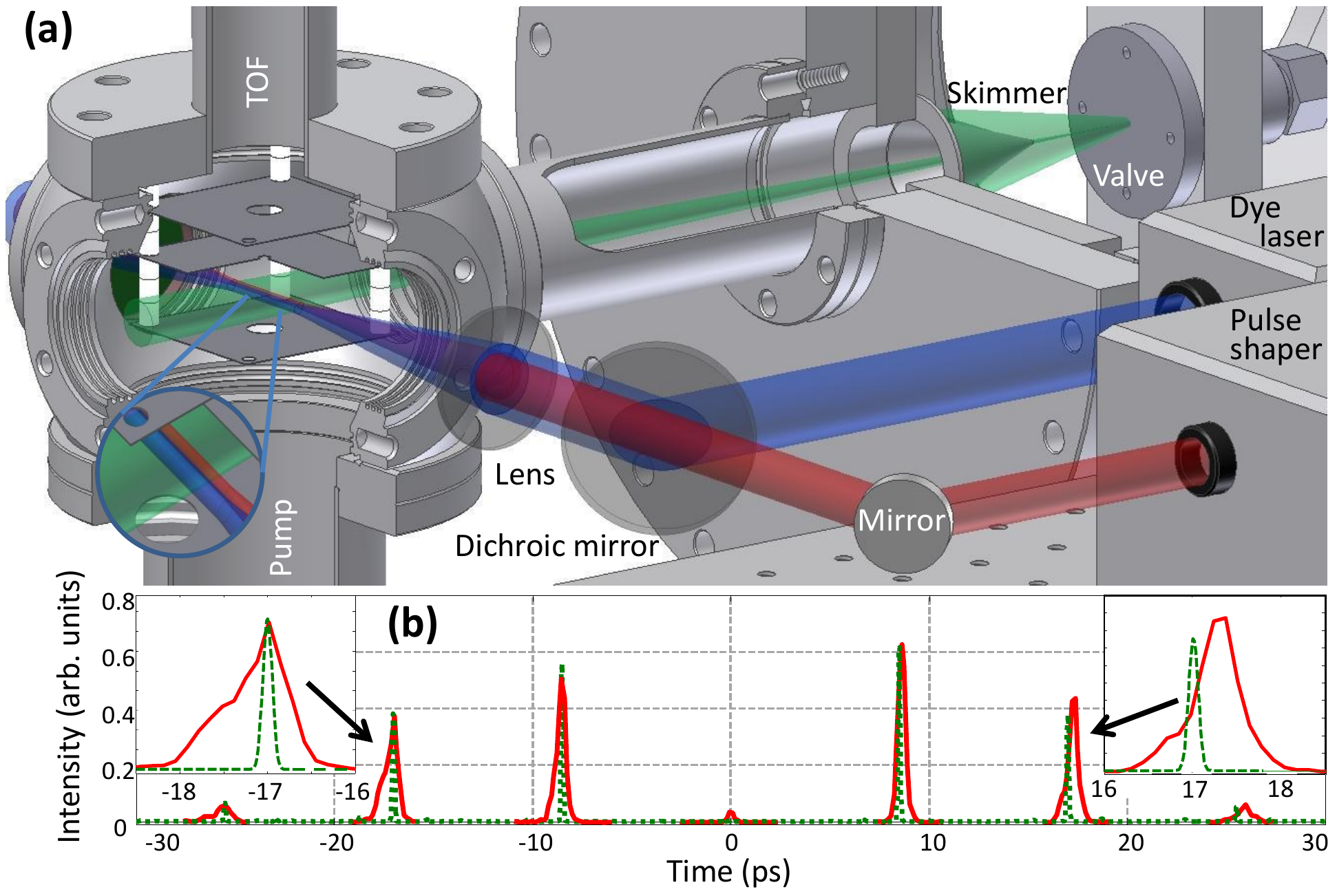}
    \caption{(Color online) (a) Experimental setup . Cold nitrogen molecules from a supersonic expansion enter the detection chamber through a 1 mm-diameter skimmer. The rotational temperature, calculated from the REMPI spectrum, is 6.3 K. The molecules are excited by a femtosecond pulse train produced with a pulse shaper (red beam) and ionized 100 ns later with a narrowband UV pulse (blue beam) shifted downstream with respect to the excitation pulse. The ions are extracted and detected with a standard time-of-flight (TOF) apparatus. (b) Example of the intensity envelope of a pulse train implemented in this work ($A=2.5$, $\tau =$ 8.5 ps). Measured (red solid line) and calculated (green dashed line) intensity profile. The insets show the discrepancies attributed to the finite optical resolution of the shaper.}
    \label{FigSetupTrain}
\end{figure}

Understanding the dynamics of molecules driven by periodic laser fields in the context of periodically kicked rotor may offer new tools for numerous schemes of coherent control of chemical reactions \cite{Zare1998, ShapiroBrumerBook}, laser cooling and trapping of molecules \cite{Forrey02}, control of molecular collisions \cite{Tilford2004}, deflection of molecular beams \cite{Purcell2009, Gershnabel2010}, high harmonic generation \cite{Itatani2005, Wagner2007} and molecular scattering from surfaces \cite{Kuipers1988, Khodorkovsky2011}. Being proportional to the molecular moment of inertia, the rotational revival time is different for different molecules, even if their chemical structures are identical as in the case of molecular isotopologues. Differences in revival times have been used to selectively align a single species in a mixture of molecular isotopologues \cite{Fleischer2006} by utilizing constructive and destructive interference in rotational excitation by a pair of short laser pulses \cite{Lee06a}. The effect of quantum resonance provides an alternative approach to selective coherent control of molecular rotation in mixtures \cite{Floss11}.

Our experimental setup is depicted in Fig.\ref{FigSetupTrain}(a). Cold nitrogen was produced by a supersonic expansion from a pulsed valve nozzle (Even-Lavie valve, EL-5-C-S.S.-2010). After entering the second chamber through a 1 mm-diameter skimmer, the molecules were excited by a femtosecond pulse train and probed by a resonance enhanced multi-photon ionization (REMPI) with a second narrowband pulse. Probe pulses were focused downstream from the excitation region and timed in such a way as to account for the molecular motion between the two spots. The ions were extracted with a time-of-flight (TOF) apparatus and the total ion signal was measured with a microchannel plate detector. The ion signal was averaged over 20 pulses.

The excitation pulse train was obtained by phase-only shaping of a linearly polarized 600 $\mu$J pulse originated from a regenerative amplifier (Spectra-Physics, Spitfire, 2 mJ at 1 KHz repetition rate). We used a home built 4$f$ Fourier pulse shaper \cite{Weiner2000} based on a 640-pixel liquid crystal spatial light modulator (CRi). To avoid strong multi-photon ionization of nitrogen by the femtosecond train, the shaper was slightly misaligned for stretching the pulses from 150 fs to about 500 fs. Sinusoidal phase modulation $\phi(\omega)=A\sin((\omega-\omega_0) \tau)$ was introduced to the input pulse spectrum to create a pulse train in the time domain, $E(t)=\sum_{n=-\infty}^{\infty}{J_n(A)\varepsilon(t+n\tau ) \cos(\omega_0t)}$, where $\varepsilon(t)$ and $E(t)$ represent the electric field envelopes of the input and output pulses, respectively, $A$ is the spectral phase modulation amplitude, $J_n(A)$ is the Bessel function of the first kind, $\tau $ is the train period and $\omega_0$ is the carrier frequency of the input field. High resolution of the shaper ($\Delta\lambda$=0.04nm/pixel) allowed us to produce pulse trains of seven pulses with a period of up to 10 ps. An example of a pulse train (both calculated and measured) with $A=2.5$ and $\tau=8.5$ ps is shown in Fig.\ref{FigSetupTrain}(b). Due to the finite resolution of the shaper, the end pulses at $\pm17, \pm$26 ps are slightly distorted, but are still shorter than the rotational period of N$_{2}$ in the highest rotational state ($J=7$) accessible with the available total pulse energy of 190 $\mu$J after the shaper. To achieve high intensity of the laser field, required for driving multiple Raman transitions between the rotational levels, pump and probe beams were focused by a 150 mm focal length lens. After measuring the focal spot size, the pump laser intensity was estimated on the order of $5\times10^{12}$ W/cm$^2$. This corresponds to the dimensionless total kick strength $P\approx7$. The latter corresponds to an average amount of angular momentum (in units of $\hbar$) transferred from the field to the molecule, and is defined as $P=\Delta\alpha/(4\hbar)\int\epsilon^2dt$, where $\Delta\alpha$ is the anisotropy of the molecular polarizability, and $\epsilon$ is the electric field strength \cite{Averbukh2001}.

\begin{figure}
\centering
    \includegraphics[width=.9\columnwidth]{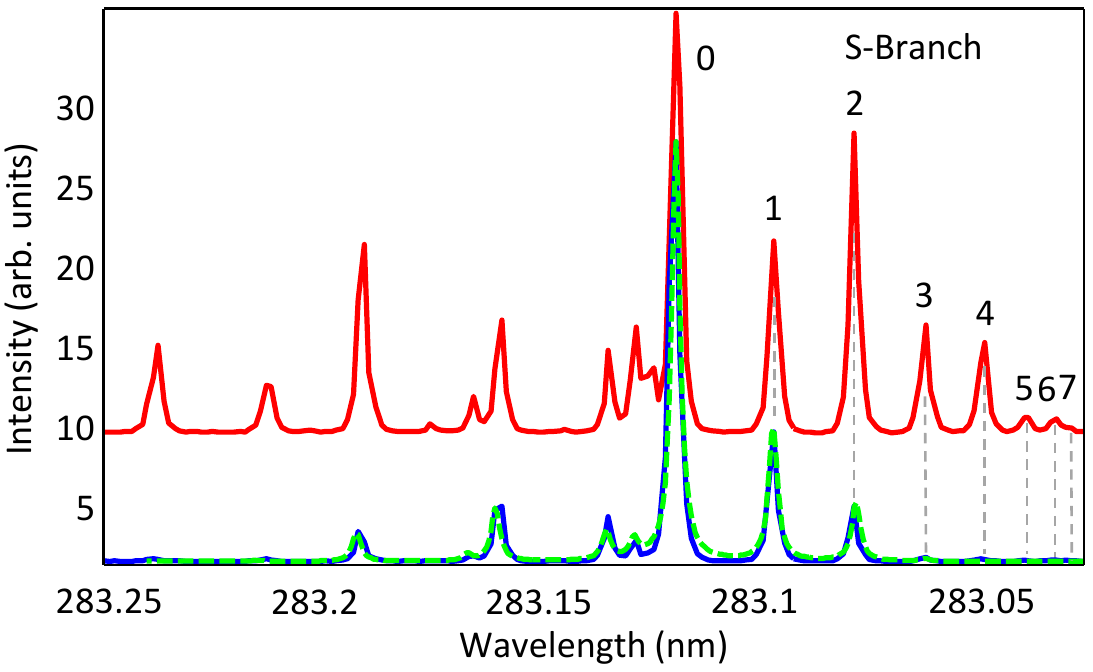}
     \caption{(Color online) REMPI spectra of $ ^{14}N_2$. Relevant peaks of the S-branch are labeled by the corresponding $N''$ numbers. Prior to the application of a femtosecond pulse train, the distribution of rotational population is thermal and corresponds to 6.3K (lower solid blue line - experiment, dashed green line - calculations \cite{HerzbergBook}). At this temperature, only $N''=0, 1$ and 2 are populated significantly. An example of the REMPI spectrum of rotationally excited molecules is shown by upper solid red line. For a total laser kick strength used in our experiments, states up to $N=7$ are populated.}
  \vskip -.1truein
  \label{FigREMPI}
\end{figure}

The rotational distribution was probed by narrowband nanosecond pulses from a tunable dye laser (Sirah, Precision Scan, 2 mJ at 283 nm and 10 Hz repetition rate). Nitrogen molecules were ionized via a ``2 +2'' resonance enhanced multi-photon ionization, with a two-photon resonant transition $a^{1}\Pi _{g}(v'=1) \leftarrow X^{1}\Sigma _{g}^{+}(v''=0)$. The frequencies of the S-branch transitions ($\Delta N=2$) are well separated, allowing us to detect the population of the first eight rotational levels $N''=0,1,...,7$ of the ground electronic state. REMPI spectrum of $^{14}N_{2}$  molecules before and after the application of the excitation laser field is shown in Fig.\ref{FigREMPI}.

\begin{figure}
\centering
    \includegraphics[width=.9\columnwidth]{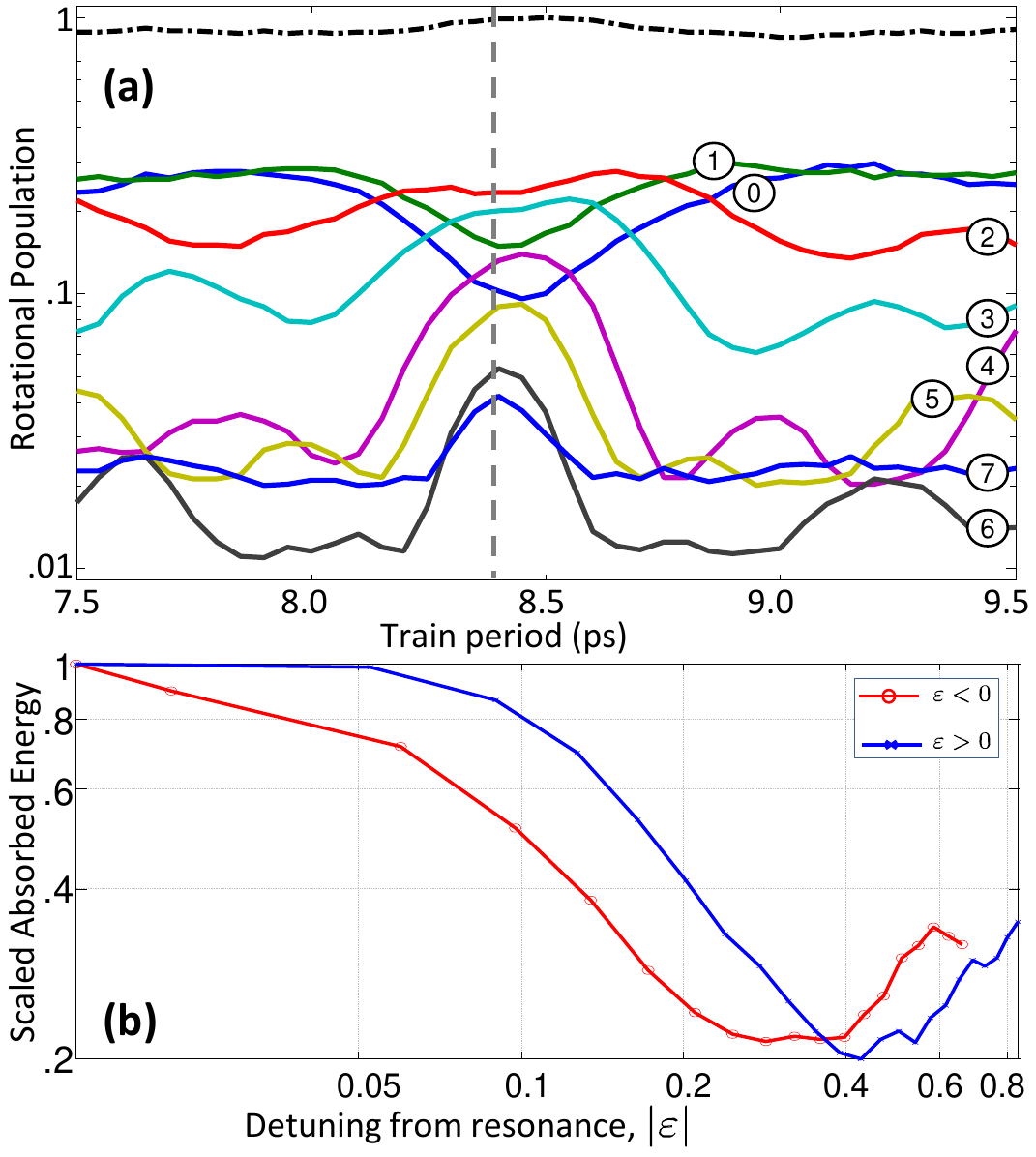}
     \caption{(Color online) (a) Relative populations $S_{N''}$ of the rotational levels of $^{14}N_{2}$ as a function of the pulse train period. Each curve is labeled with the corresponding rotational quantum number $0\leq N'' \leq 7$. Vertical dashed line marks the rotational revival time, at which the efficiency of the rotational excitation by a periodic train of pulses is significantly enhanced. Total population, $\sum_{N''}{S_{N''}}$, is depicted by the upper dash-dotted line. (b) Total absorbed energy, normalized by its resonant value, as a function of the dimensionless detuning from quantum resonance (see text).}
  \vskip -.1truein
  \label{FigN14}
\end{figure}

To measure the total rotational energy absorbed by the molecules, we detected the transfer of population from the initially populated ($N''=0,1,2$) to higher ($N'' \leq 7$) rotational states as a function of the pulse train period. REMPI signal was measured by tuning the probe wavelength to the corresponding peak of the REMPI spectrum, and recording the ion signal while scanning the train period with the pulse shaper. To determine the relative population of each rotational level, $S_{N''}$, REMPI signals were scaled with the corresponding two-photon line strength factors for the $N'\leftarrow N''$ rotational transitions and the nuclear spin degeneracy weights \cite{Aoiz99}. The results are shown on the logarithmic scale in Fig.\ref{FigN14}(a). One can see that the highest increase of rotational energy (i.e. the most efficient population transfer up the rotational ladder) occurs around 8.4 ps - the revival time of $ ^{14}N_2$, where the population is efficiently transferred from the lower to higher angular momentum states, as indicated by the dips and peaks in $S_{0,1}$ and $S_{2,...,7}$, respectively. Small drop of the total population ($\sum_{N''}{S_{N''}}$) below unity away from quantum resonance can be attributed to the resonance-enhanced anisotropy of the spatial distribution of the molecular angular momentum, which was not taken into account in our conversion of the REMPI signal to the relative populations. In panel (b), the population distribution is converted to the scaled total energy absorbed by the molecules and plotted as a function of the dimensionless detuning from quantum resonance, $\varepsilon=2\pi[t/T_\text{rev}-1]$. Both plots, for $\epsilon<0$ and $\epsilon>0$, resemble well the characteristic oscillating behavior predicted in \cite{Floss11}.

\begin{figure*}
\begin{center}
    \begin{minipage}[t]{1\linewidth}
    \includegraphics[width=.9\columnwidth]{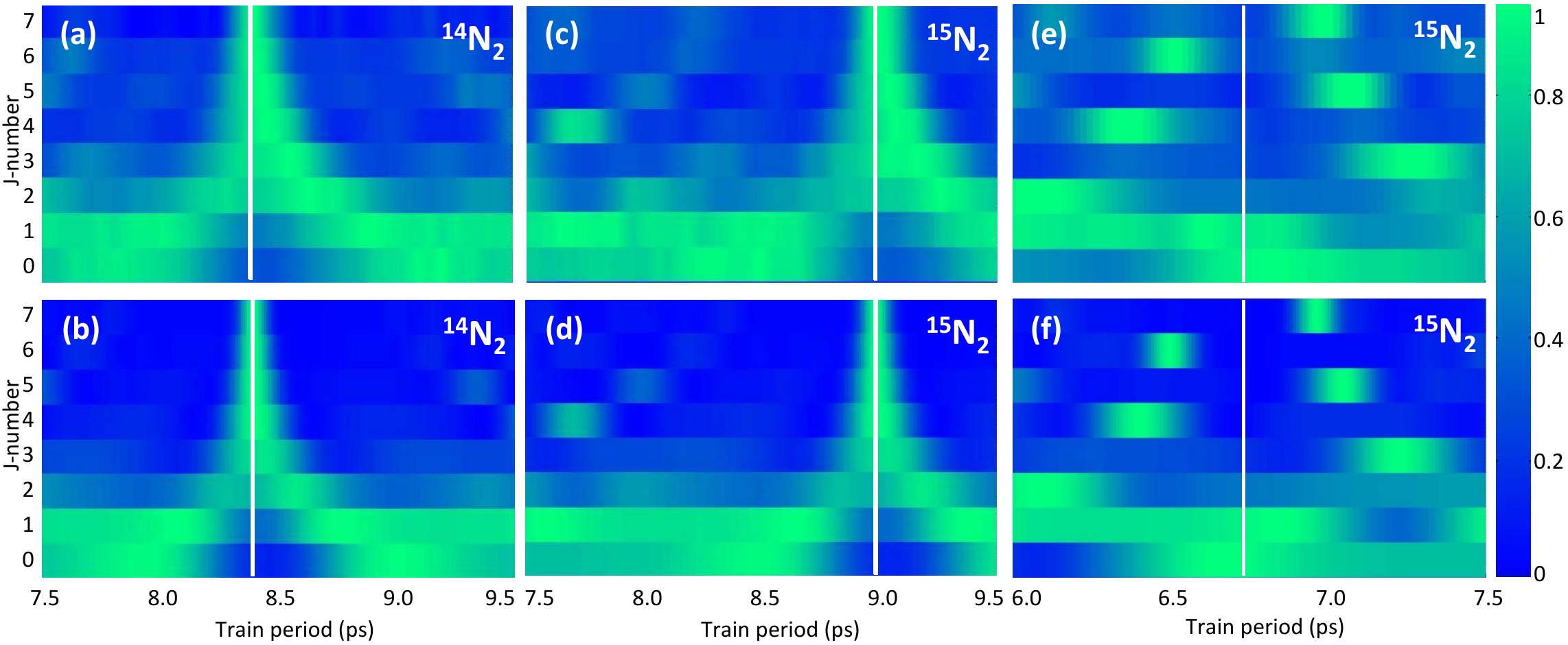}
    \caption{(Color online) Experimental results (upper row) and numerical simulations (bottom row) of a normalized REMPI signal (color coded) for different $N''$ states as a function of a pulse train period. Left and middle columns correspond to $^{14}N_2$ and $ ^{15}N_2$, respectively. For a pulse train period equal to the revival time (white vertical lines) the population is efficiently transferred from the initial states $N''=0,1,2$ to higher states $N''=3, 4,...7$. Right column shows the rotational population distribution for $^{15}N_2$ as a function of the pulse train period around $\nicefrac{3}{4}T_{\text{rev}}$.}
    \label{FigResults}
    \end{minipage}\hfill
  \end{center}
\end{figure*}

To emphasize the resonant nature of the rotational energy accumulation by the molecules, we normalized the ion signal for each $N''$ separately (i.e. divided each curve in Fig.\ref{FigN14}(a) by its maximum value) and plotted the results as a two-dimensional map in Fig.\ref{FigResults}(a). The effect of quantum resonance is clearly demonstrated by the significantly enhanced population transfer to higher $J$-states in the case when the period of the femtosecond pulse train matches the rotational revival time. Remarkably, even for a relatively small number of pulses in the train (see Fig.\ref{FigSetupTrain}), the observed resonance is quite narrow and can be used for selective rotational excitation of molecular mixtures. We demonstrate this by performing the same measurement with two isotopologues of nitrogen, which have identical chemical structure but different rotational properties, reflected by the difference in the revival time.  The detected population transfer efficiency for $^{14}N_{2}$ ($T_{\text{rev}}=8.38$ ps) and $^{15}N_{2}$ ($T_{\text{rev}}=8.98$ ps) is shown in Figs.\ref{FigResults}(a) and (c), respectively. By tuning the pulse train period to 8.4 or 9.0 ps, one can induce the rotation of a selected isotopologue, while keeping the other in its initial rotational state.

The results of numerical calculation of the rotational population transfer for both isotopologues are shown in the bottom row of Figure \ref{FigResults} (panels (b) and (d)). In our simulations, we averaged over the fast oscillations of the electric field, expanded the wave function in the spherical harmonics, and solved numerically the time-dependent Schr\"odinger equation to obtain the expansion coefficients \cite{Floss11}. The actual shape of the generated pulse trains has been included in the calculations. To take into account thermal effects, we averaged over the initial states, where each result is weighted by the Boltzmann coefficient (including nuclear spin statistics) of the initial state.

Molecular spin isomers, such as para- and ortho-isomers of $ ^{15}N_2$, exhibit identical revival times yet different structure of rotational levels. Para-nitrogen (total nuclear spin $I=0$) does not have odd $N$ states in its rotational spectrum, whereas even $N$'s are missing in the spectrum of ortho-nitrogen (total nuclear spin $I=1$). In this case, coherent control of molecular rotation can be based on fractional, rather than full,  rotational revivals. An example is shown in Figure \ref{FigResults}(e) and (f) (experiment and theory, respectively), where selective population transfer into even (odd) excited rotational states of para (ortho) nitrogen is achieved by tuning the period of the femtosecond pulse train slightly below (above) $\nicefrac{3}{4} T_{\text{rev}}$. We note that unlike the selectivity of molecular alignment demonstrated with a sequence of two pulses \cite{Renard2004, Fleischer2007}, our results suggest that pulse trains may offer a new way of \textit{state selective} rotational excitation. With a relatively small number of pulses used here, the selectivity is limited to low lying $N''$ states (e.g. $N''=4$ at $t\approx 6.3$ ps or $N''=5$ at $t\approx 7.1$ ps).

In summary, periodically kicked quantum rotor was studied experimentally and theoretically, using nitrogen molecules and a train of ultrashort laser pulses. Quantum nonlinear resonances have been demonstrated in the efficiency of rotational excitation as a function of the pulse train period. Enhancement of the rotational energy transfer has been observed at and around full and fractional revivals of the molecular quantum wavepacket. The ability to utilize quantum resonances for the selective excitation of different molecular isotopologues and nuclear spin isomers has been shown.

\begin{acknowledgements}
This work has been supported by the CFI, BCKDF and NSERC. IA and JF acknowledge support from the ISF and DFG (LE 2138/2-1). SZ is a recipient of an Alexander Graham Bell scholarship from NSERC. JF is a recipient of a fellowship from Minerva Foundation.
\end{acknowledgements}


\end{document}